# Effect of wall slip on laminar flow past a circular cylinder

Yan-cheng Li[1,2,3], Sai Peng[4*] and Taiba Kouser[5]

[1]Key Laboratory of High Performance Ship Technology, Wuhan University of Technology, Ministry of Education, 430063, Wuhan, China

[2]School of Naval Architecture, Ocean and Energy Power Engineering, Wuhan University of Technology, 430063, Wuhan, China

[3]School of Naval Engineering, Wuxi Institute of Communications Technology, 214151, Wuxi, China

[4]Shenzhen Key Laboratory of Complex Aerospace Flows, Department of Mechanics and Aerospace Engineering, Southern University of Science and Technology, 518055, Shenzhen, China

[5]Department of Mathematics, Government College University Faisalabad, 38000, Pakistan

**Abstract**

A numerical study of two-dimensional flow past a confined circular cylinder with slip wall is performed. A dimensionless number, Knudsen number ($Kn$) is used to describe the slip length of cylinder wall. The Reynolds number ($Re$) and Knudsen number ($Kn$) ranges considered are $Re = [1, 180]$ and $Kn = [0, \infty)$, respectively. Time-averaged flow separation angle ($\bar{\theta}_s$), dimensionless recirculation length ($\bar{L}_s$) and tangential velocity ($\bar{u}_\tau$) distributed on the cylinder's wall, drag coefficient ($\bar{C}_d$) and drag reduction ($DR$) are investigated. The time-averaged tangential velocity distributed on the cylinder's wall fits well with the formula

$$\bar{u}_\tau = U_\infty \cdot \left[ \frac{\alpha}{1+\beta e^{-\gamma(\pi-\theta)}} + \delta \right] \cdot \sin\theta$$, where the coefficients ($\alpha$, $\beta$, $\gamma$, $\delta$) are related to $Re$ and $Kn$, and $U_\infty$ is the incoming velocity. Several scaling laws are found, $\log(\overline{u_{\tau max}}) \sim \log(Re)$ and $\overline{u_{\tau max}} \sim Kn$ for low $Kn$ ($\overline{u_{\tau max}}$ is the maximum tangential velocity on the cylinder's wall), $\log(DR) \sim \log(Re)$ ($Re \leq 45$ and $Kn \leq 0.1$) and $\log(DR) \sim \log(Kn)$ ($Kn \leq 0.05$). At low $Re$, $DR_v$ (the friction drag reduction) is the main source of $DR$. However, $DR_p$ (the differential pressure drag reduction) contributes the most to $DR$ at high $Re$ ($Re > \sim 60$) and $Kn$ over a critical number. $DR_v$ is found almost independent to $Re$.

**Keywords:** circular cylinder, drag reduction, slip length, numerical simulation.

## 1. Introduction

Flow past a cylinder is an important research topic in the field of fluid mechanics and hydraulic engineering, and it is also one of the most challenging problems for complex flow patterns. For the simplest case of viscous flow past a confined circular cylinder, depending on the value of the Reynolds number ($Re$), the flow experiences several transitions from one flow regime to another [1-3]. At very low Reynolds numbers ($Re<<1$), since fluid inertia is neglected, fluid elements are able to adjust the shape of the submerged blunt body, so as to closely follow its contour, that is, the flow keeps attached to the surface. With the gradual increase of $Re$, fluid inertia increases and reaches a point where the fluid inertia leads to the formation of a reverse pressure gradient at a certain point on the surface of the bluff-body. This causes the flow to detach and thereby leads to the appearance of a separation bubble at $Re \approx 5$. By increasing $Re$, the separation bubble grows in size until the wake becomes asymmetric (about the mid-plane) and finally, it becomes unstable at $Re \approx 48$. Beyond this $Re$, the vortices are shedded and the flow field becomes periodic in time. At $Re \approx 180$, the overall flow field exhibits some features of three-dimensional flow. Therefore, the complexity of this seemingly simple flow increases with increasing $Re$. In addition, a large number of engineering models can be simplified to flow through the main body of the cylinder, such as wind blows through high-rise buildings and sea water flowing through submarine pipeline [1-3].

*Corresponding Auther: Sai Peng (273986081@qq.com)

As for the boundary condition of the solid-liquid interface, the non-slip boundary condition has been widely recognized, and its rationality has been proved by experimental research. Due to the rapid development of current measurement technology, a lot of experimental work has been devoted to analyze the behavior of micro-scale and nano-scale fluids recently[4,5]. The continuum hypothesis is still applicable to this field, and the Navier-Stokes equations can still be used to predict the fluid motion, although the effectiveness of the non-slip boundary condition has been proved by experiments to break the extremely limited systems[6-8]. A slip boundary is usually characterized by a slip length, which is the fictitious distance between the physical interface and an imaginary surface with zero tangential velocity in the body [9]. For example, in the flow through thin microchannels [10], the apparent slip length was directly checked on hydrophobic surface. Superhydrophobic surfaces were used to achieve large slip lengths up to 400μm, and the liquid surfaces are mostly in contact with air that is trapped in the structured or unstructured crevices of the engineered surfaces [11-13]. The practice of using superhydrophobic surface to create a partial slip boundary condition has been confirmed in the literature, and can reduce the drag in laminar [14,15] and turbulent flows [16-18].

In particular, wall slip has a special effect on the flow past a circular cylinder. Numerical [19,20] and experimental studies [14,19, 20] indicate that the wall slip modifies the flow field, such as weakening the flow fluctuation, shortening the wake zone and increasing the frequency of vortex shedding, etc. First, the related experimental research is reviewed. The influence of partial slip conditions with different superhydrophobic surfaces when $Re$ is as high as 10, 000 is studied by Muralidhar *et al.* [21]. They noticed that the flow behavior depends upon the arrangement of ridges on the superhydrophobic surface. Compared with a smooth cylinder, the frequency of shedding increases. Compared with those arranged in the normal direction, the ridges arranged in the flow direction show a higher shedding frequency. They also noticed a delayed onset of vortex shedding and an elongation of the recirculation bubble for a cylinder with a superhydrophobic surface. Coating superhydrophobic sand of size varying from 50 to 710 μm on the cylinder to increase surface roughness, Brennan *et al.* [18] obtained the maximum drag reduction up to 28% in the $Re$ ranges from 10, 000 to 40, 000. Trough sanding sandpapers with two different grit sizes into streamwise and spanwise directions, respectively, on the hydrophobic cylinder surfaces, Kim,

Kim & Park [16] examined the effect of the slip direction on the flow field and drag coefficient. The results demonstrated that the rough hydrophobic surface enhanced the turbulence in the flow above the circular cylinder and along the separating shear layers, resulting in a delay of the flow separation and early vortex roll-up in the wake, and reduction in the size of the recirculation bubble in the wake up to 40%, while the drag reduction of less than 10% was estimated from a wake survey for *Re* from 0.7-2.4×10$^4$. The effect of superhydrophobicity with *Re* ranging from 1300 to 2300 on vortex-induced motion of cylinder was studied by Daniello *et al.* [22]. They found that slippage would reduce the root mean square lift and the amplitude of the oscillating cylinder. They also observed that wall slip increases the size (length and width) of the recirculation bubble while decrease the intensity of shed vortices and the lift coefficient.

Second, the related numerical research is reviewed. 2D-DNS (Direct numerical simulation) [19] was applied to derive the critical Knudsen number (*Kn*) for the disappearance of wake recirculation and trailing vortex-shedding of the flow over a circular cylinder with Reynolds number (*Re*) less than 800. Seo *et al.* [23] investigated the flow parameters via numerical simulation of shallow-water flow over a cylinder, and found that the wall slip reduced the wall vortex and the shear layer developed along the cylinder surface by the Navier-slip condition tended to push the separation point away to the rear stagnation point. Li *et al.* [24] studied the slip distribution on the flow over the circular cylinder and concluded that the flank slip had the biggest drag reduction effect compared with the slip configurations for the fore - side slip and aft-slip for *Re* = 100~500 and *Kn* = 0.2. D'Alessio *et al.* [25-26] proposed two methods to solve the two-dimensional problem of the unsteady flow of a viscous incompressible fluid past a circular cylinder or elliptic cylinder subject to impermeability and Navier-slip conditions on the surface. The first takes the form of a double series solution where an expansion is carried out in powers of the time, *t*, and in powers of the parameter $\lambda = \sqrt{8t/Re}$. This approximate analytical solution is valid for small times following the start of the motion and for large Reynolds numbers. The second method involves the spectral finite difference program, which is used to numerically integrating the complete Navier-Stokes equations expressed by the flow function and vorticity. They proved that the two methods of solution are in excellent agreement for small times and moderately large *Re*. Then, using these two methods, they studied flow past a slippery circular cylinder and elliptic cylinder. They found a drag reduction and a suppression in vortex shedding for this flow. Ren e*t al.* [27] numerically studied wall partial slip on a rotating circular cylinder. The partial slip was considered as alternating shear-free and no-slip boundary conditions. Gas-fraction (*GF*) could be used to evaluate wall partial slip for this special configuration. They considered the non-dimensional rotation rates of $0 \leq \alpha \leq 6$ at *Re* = 100. They found wall slip play different roles in different flow regimes so that the first critical rotation rate decreases monotonically with the increase of *GF*, while the second and the third critical rotation rates increase monotonically with increasing *GF*. Some of the previous work involves three-dimensional situations. Kouser *et al.* [28] numerically studied wall slip on three-dimensional NACA0012 hydrofoil. Surface heterogeneity is taken into account by periodic no-slip and shear-free grates with ridges arranged normal to the inlet flow direction. The study is conducted for a range of $0^0$ to $25^0$ degree angles of attack and at a fixed Reynolds number of 1000. Bubble separation is delayed for higher gas-fraction (*GF*), and the vortex shedding effects is minimized. As a result, flow remains two-dimensional (*2D*) for higher angles of attack as compared to flow over three-dimensional (3*D*) hydrofoil with no-slip boundary. Mode C with smaller wavelength is observed at $\alpha$ = $15^0$. You & Moin [29] presented a numerical study, using direct numerical simulations (DNS) and large eddy simulations (LES), for Reynolds numbers of 300 and 3900. They observed that a hydrophobic treatment on a microscale circular cylinder leads to reduction in the mean drag and the root mean square lift coefficient values. In addition, the drag reduction in the laminar vortex shedding regime is mainly due to the reduction of surface friction, while the reduction of drag in the shear layer transition regime is due to a delay in separation.

In recent years, previous researches have paid more attention to the effect of wall slip or slip distribution on the statistical parameters, such as drag coefficient, lift coefficient and Strouhal number, flow regime (whether the flow is separated or not, steady or unsteady, two-dimensional or three-dimensional modes) using experimental and numerical simulation methods. In all, the influence of wall slip on the statistical parameters and flow regime is due to the specific velocity distribution on the wall. Li *et al.* [24] gives an analytical solution through the matched-asymptotic expansion. This analytical solution is suitable for only low Reynolds numbers (*Re*<<1). Kumar *et al.* [30] gives an asymptotic theory for the high-Reynolds-number flow past a shear-free circular cylinder ($Re \to \infty$). However, between these two extreme situations, there is not enough quantitative report about the velocity distribution on the wall.

It is well known that wall slip can reduce the drag of the cylinder. However, there is a lack of quantitative understanding of the influence of *Re* and *Kn* on drag reduction. For fully developed channel flow, wall slip on channel drag reduction is only the result of friction drag reduction. However, the drag acting on cylinder is composed of two parts, pressure difference drag and friction drag. At low *Re* (Stokes flow), both of them account for about 50%. But at high *Re*, pressure difference drag occupies most of the total drag. Maghsoudi *et al.* [31] founds the differential pressure drag is enhanced at low *Re* when slip velocity exits. It is still unknown which part of the drag is reduced by the wall slip at different Reynolds numbers.

The aim of this study is to obtain the velocity distribution on the wall, the drag reduction and the contribution to drag reduction for laminar flow past a circular cylinder through numerical calculation. In this paper, *Re* is set as from 1 to 180. Beyond *Re* = 188 [32], the flow exhibits mode A three-dimensional transition for no-slip case. In order to save computing resources, the calculation is set in two dimensions in this paper, so the maximum *Re* considered is 180. This article is organized as follows: Section 2 presents the geometry and methodology used. In section 3, two serial results are compared with the literature to check accuracy of our code used. Section 4 illustrated the the flow characteristic parameters such as time-averaged separation angle, dimensionless recirculation length and the velocity distributions on the wall, drag coefficient and drag reduction are investigated. In the last Section, a brief conclusion is given.

2. **Problem formulation and theoretical methods**

As shown in figure 1(a), a two-dimensional flow past a cylinder with diameter *D* placed in an infinite domain at a uniform speed $U_\infty$ = 1 is simulated. The length and height of the partial domain are $L_u + L_d$ and $2L_y$, respectively. The cylinder is placed at a distance $L_u$ from the inlet boundary and a distance $L_d$ (downstream length) from the outlet boundary. The governing equations have been rendered dimensionless using the flowing scaling variables: $D$ for length variable, including the slip length, $U_\infty$ for velocity, $D/U_\infty$ for time, $\rho U_\infty^2$ for pressure, and $U_\infty/D$ for viscosity, respectively. For the unsteady two-dimensional incompressible flow, the dimensionless governing equations are given as:

$$\nabla \cdot \boldsymbol{u} = 0,$$
$$Re \cdot \left(\frac{\partial \boldsymbol{u}}{\partial t} + \boldsymbol{u} \cdot \nabla \boldsymbol{u}\right) = -\nabla p + 2\nabla \cdot \boldsymbol{S}, \quad (1)$$

where $\boldsymbol{S} = (\nabla \boldsymbol{u} + (\nabla \boldsymbol{u})^T)/2$ denotes the rate-of-strain tensor, $Re = \rho U_\infty D/\mu$ is Reynolds number and $\mu$ is viscosity.

The slip boundary conditions are as follows[19]:

$$\boldsymbol{n} \cdot \boldsymbol{u} = 0,$$
$$\boldsymbol{n} \times \boldsymbol{u} = Kn \cdot D \cdot \boldsymbol{n} \times (\boldsymbol{S} \cdot \boldsymbol{n}), \quad (2)$$

where, $Kn = \lambda/D$ is Knudsen number and $\lambda$ is the slip length and $\boldsymbol{n}$ is the exterior normal vector direction. The other boundary conditions are set as: constant velocity at the inlet ($\boldsymbol{u} = [1,0]$), symmetry boundary at upper and down boundary, and a zero-pressure at outlet boundary($p_\infty = 0$).

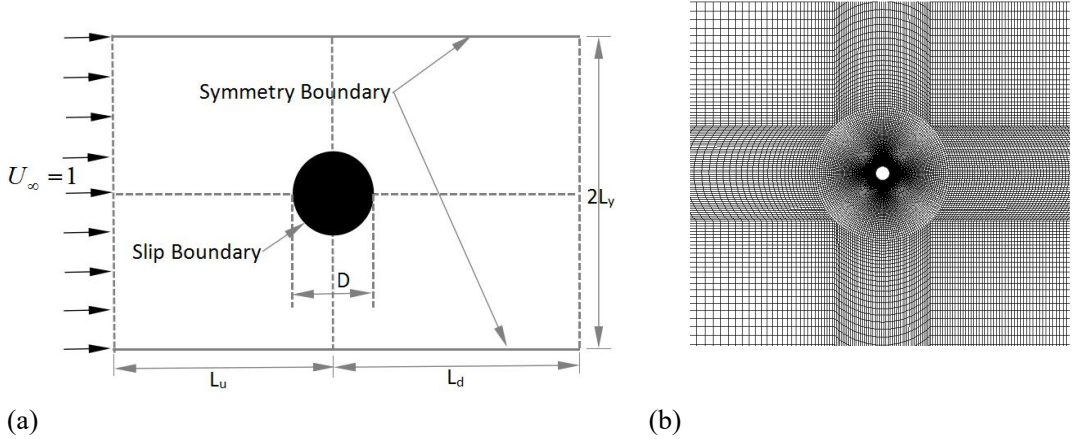

(a) (b)

**Figure 1.** (a) Schematics of the flow over a confined circular cylinder in a free domain; (b) Computational grid near the cylindrical wall for $Re$ from 1 to 45.

Figure 2 shows a schematic diagram of the angle $\theta$ and the wall slip velocity $u_\tau$, where $\theta$ starts from the front end of the cylinder. The slip tangential velocity on the wall can be obtained by the formula,

$$u_\tau = u \cdot \sin\theta + v \cdot \cos\theta, \quad (3)$$

in Cartesian coordinate system. The pressure coefficient ($C_p$) is defined by the expression:

$$C_p(\theta) = \frac{2(p(\theta) - p_\infty)}{\rho U_\infty^2}, \quad (4)$$

where $p(\theta)$ is the surface pressure at $\theta$ and $p_\infty$ is the pressure at the outlet boundary. The total drag coefficient $C_d$, that is, the sum of the friction drag coefficient and the differential pressure coefficient, can be written as follows,

$$C_d = \frac{2F_d}{\rho U_\infty^2 D},  \tag{5}$$

where $F_d$ is the total drag force acting on the cylinder. The friction and differential pressure drag coefficients $C_{dv}$ and $C_{dp}$ are obtained by integrating on the cylindrical surface as follows,

$$C_{dv} = \frac{2\int_0^{2\pi} \tau_{wx} \sin\theta d\theta}{\rho U_\infty^2},  \tag{6}$$

and

$$C_{dp} = \int_0^{2\pi} C_p \cos\theta d\theta,  \tag{7}$$

respectively, where $\tau_{wx}$ is shear stress component in $x$ direction on of the cylinder surface.

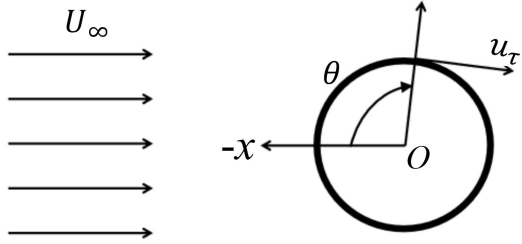

**Figure 2.** Diagram of angle and wall slip velocity.

Numerical simulation is performed using the commercial software FLUENT 15.0. The detailed descriptions of the numerical solution procedure are available elsewhere [33-35], and only the salient features are recapitulated here. Briefly speaking, the structured quadrilateral cells of uniform and non-uniform grid spacing are generated using the commercial grid tool ICEM. The two-dimensional laminar segregated solver is used to solve the incompressible flow on the collocated grid arrangement. Besides, the QUICK scheme is applied to disperse the convective terms in the momentum equations, and the pressure-velocity coupling is solved by the coupled scheme. The time terms are discretized by second-order implicit scheme. We resolve the system of algebraic equations with the Gauss-Siedel(G-S) point-by-point iterative method. The slip wall boundary conditions are achieved by loading wall tangential force through FLUENT UDF functions. The equation (2) could be rewritten as follows:

$$u_\tau = \lambda \frac{du_\tau}{dn}.  \tag{8}$$

Moreover,

$$\tau_w = \mu \frac{du_\tau}{dn}.  \tag{9}$$

$\tau_w$ is shear stress on the cylinder surface. From equation (8) and equation (9), we can get

$$\tau_w = \mu \frac{u_\tau}{\lambda}.  \tag{10}$$

Written equation (10) in the form of components in Cartesian coordinate system,

$$\tau_{wx} = \mu \frac{u}{\lambda} \quad \text{and} \quad \tau_{wy} = \mu \frac{v}{\lambda}. \tag{11}$$

The wall shear stress components are loaded on FLUENT software by UDF functions.

In this study, simulated $Re$ ranges from 1 to 180. For low $Re$, such as 1, it needs large computational spatial domain to devoid the blockage effect on flow parameters [36]. Therefore, we use a large-scale computational spacial region set as $2L_y = 250D$, $L_u = 250D$ and $L_d = 750D$, relatively, as shown in figure 1(a) for $Re$ from 1 to 45. The grid near the cylindrical wall is shown in figure 1(b). The computational spacial domain is discredited with O-type structural Grids, and 196 nodes are uniformly arranged on the wall with the innermost mesh thickness $\delta = 0.0025D$. However, for $Re$ from 45 to 180, it is not necessary to use such a large computational domain. Another computing spacial domain and the corresponding computing grid are adopted. The computational spacial region is set as $2L_y = 50D$, $L_u = 25D$ and $L_d = 75D$, relatively. The spacial computational domain is also discredited with O-type Structural Grids, and 300 nodes are uniformly arranged on the wall with the innermost mesh thickness $\delta = 0.000625D$. When $Re$ is less than 45, steady-state calculation is used. However, when $Re$ is larger than 45, unsteady calculation is used to determine the critical $Kn^{**}$ of transition from unsteady regime to steady regime. Once $Kn^{**}$ is determined, steady calculation is used for $Kn$ larger than $Kn^{**}$. For unsteady calculation, the time-step adopted is $0.025D/U_\infty$. In all our this simulation, time-averaged flow field calculation counts more than 20 time periods after the flow reaches the final state. The time-averaged variable is expressed by putting a bar over the corresponding symbol.

## 3. Validation

In order to confirm the accuracy of our simulation results, two sets of comparisons are given. First is flow over a no-slip circular cylinder. The drag coefficient, pressure coefficient, recirculation length and separation angle are compared with those in the literature. The length of recirculation bubble $L_w$ is the dimensionless distance measured from the rear of the cylinder ($\theta = \pi$) to the attachment point in the near closed streamline ($u = v = 0$) on the line of flow midline as shown in the illustration of figure 4(a). The angle of separation $\theta_s$ is the value between the leading stagnation point ($\theta = 0^0$) and the onset of flow separation from the solid surface, as shown in the illustration of figure 4(b). Second is flow over a slip circular cylinder. The transitional $Kn$ ($Kn^{**}$) for different $Re$ are considered and compared with the results of Legend et al. [19].

Table 1 summarizes and compares our and other groups results, including $\overline{C_{dp}}$, $\overline{C_{dv}}$, $\overline{C_d}$, $\overline{C_p(0)}$, $-\overline{C_p(\pi)}$ (base pressure coefficient), $\overline{L_w}$ and $\overline{\theta_s}$, for no-slip circular cylinder. It is found that the present calculated results are reasonable, since which are within 2% of the previous values. For example, $C_d$ is 2.0057 in present study, while Fornberg [37] got it as 2.0000 for $Re = 20$. $\overline{C_d}$ is 1.3389 in current study, while Park et al. [38] got is as 1.33, Xiong et al. [39] gives it as 1.34 for $Re = 100$.

The critical Reynolds number ($Re_c$) for flow separation transition (downstream recirculation bubble occurs) is slightly less than 7 of the flow past a no-slip circular cylinder [19], while our $Re_c$ is 6.5, in excellence with previous results. The critical $Kn^*$ for vanishing recirculation behind the cylinder for slip cylinder is shown in figure 3(a). Our results are within the range of less than 3% difference relatively compared with the results calculated by Legendre et al. [19]. The critical $Kn$ is approximately linear with *Re* for *Re* range from 6.5 to 40 as shown in the illustration of figure 3(a). Another transition is wake vortex shedding behind a circular cylinder. For no-slip circular cylinder, the critical Reynolds number ($Re^{**}$) is about 47.5, which coincides well with Legendre *et al.* [19] (47.5). Wall slippage delays this transition. The transition critical $Kn^{**}$ for different *Re* is plotted figure 3(b). Our results are in less than 5% difference relatively compared to the results calculated by Legendre *et al.* [19].

Table 1. Comparison of the present results with the literature values for no-slip walls.

| Source | $\overline{C_{dp}}$ | $\overline{C_{dv}}$ | $\overline{C_d}$ | $\overline{C_p(0)}$ | $-\overline{C_p(\pi)}$ | $\overline{L_w}$ | $\overline{\theta_s}$ |
|---|---|---|---|---|---|---|---|
| *Re* = 20 | | | | | | | |
| Present results | 1.2037 | 0.8020 | 2.0057 | 1.2640 | 0.5438 | 0.8973 | 136.57 |
| Fornberg [37] | - | - | 2.0000 | 1.2800 | 0.5400 | 0.9100 | - |
| D'Alessio and Dennis [40] | - | - | 1.8410 | 1.2640 | 0.5360 | 0.8750 | 136.90 |
| Park *et al.* [38] | 1.2100 | 0.8000 | 2.0100 | - | - | - | - |
| *Re* = 40 | | | | | | | |
| Present results | 0.9809 | 0.5207 | 1.5016 | 1.1416 | 0.4775 | 2.1910 | 126.59 |
| Fornberg [37] | - | - | 1.4980 | 1.1440 | 0.4600 | 2.2400 | 126.20 |
| Park *et al.* [38] | 0.9900 | 0.5200 | 1.5100 | - | - | - | - |
| *Re* = 100 | | | | | | | |
| Present results | 0.9963 | 0.3426 | 1.3389 | 1.0983 | -0.6933 | 1.4224 | 116.96 |
| Park *et al.* [38] | 0.99 | 0.34 | 1.33 | 1.0270 | -0.7311 | 1.4050 | 117.19 |
| Xiong *et al.* [39] | | | 1.34 | | | | |

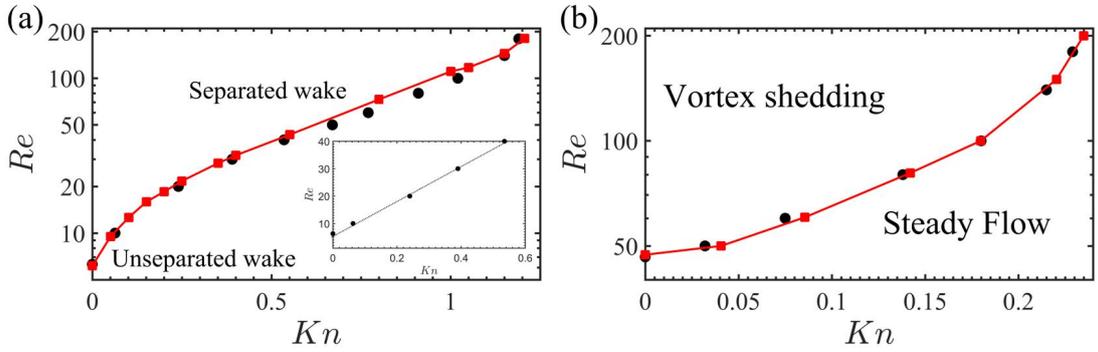

**Figure 3.** The critical Knudsen numbers ($Kn^*$ and $Kn^{**}$) for (a) downstream recirculation disappearing (b) vortex shedding disappearing at each Reynolds numbers compared with Legendre *et al.* [19]. The black circle points are present results, while the red square dots connecting into a line are results of Legendre *et al.* [19].

## 4. Results and Discussion

### 4.1 Recirculation length and separation angle

For flow past a blunt body, recirculation and separation are one of the most important characteristics for this flow pattern, which is often quantified in terms of $L_s$ and $\theta_s$. Through numerical simulation [36,38,41,42] and experimental research [42,43], the length and separation angle of the downstream recirculation bubble have been widely studied in previous literature. The popular opinion is that $L_w$ is linearly related with $Re$, and the separation angle is linear with $Re^{-1/2}$ for $10 \leq Re \leq 40$ [36,38,41,42]. At this $Re$ range, we also get similar results, as shown in figure 4. However, when $Re$ is over $Re^{**}$, such relation is not applicable. With the increase of $Re$, the length of recirculation bubble ($\overline{L}_s$) of time-averaged flow becomes shorter. Our results coincide well with the simulation reports of Park *et al.* [38] as shown in figure 4(a). When $Re$ is over 40, the relation between $\overline{\theta}_s$ and $Re$ satisfies the following equation [42],

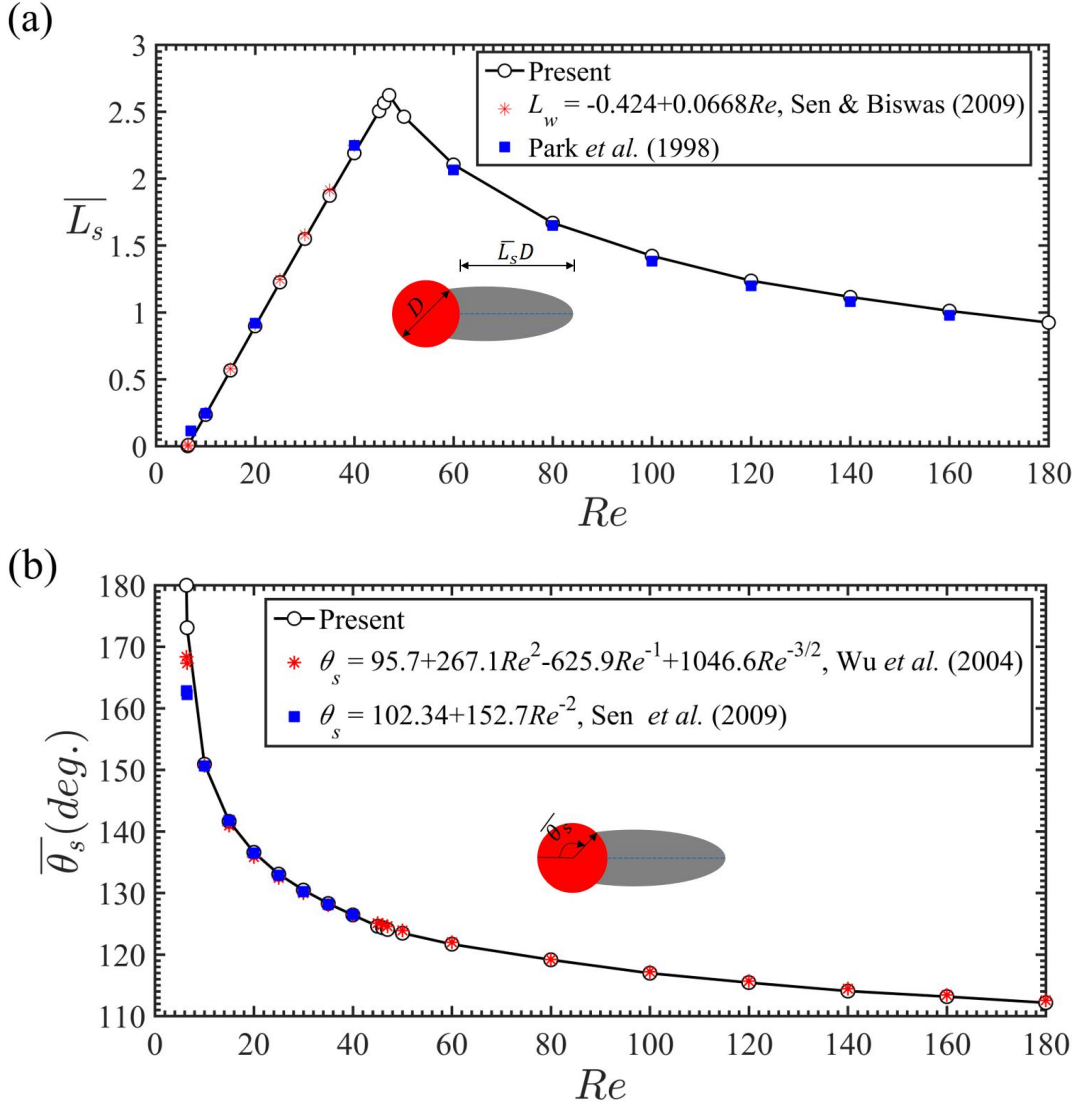

**Figure 4.** The relationship between (a) downstream recirculation bubble length ($\overline{L}_s$), (b) flow separation angle ($\overline{\theta}_s$) and Reynolds number ($Re$).

$$\overline{\theta}_s = 95.7 + 267.1 Re^2 - 625.9 Re^{-1} + 1046.6 Re^{-3/2}. \quad (12)$$

Our simulation results satisfy equation (12). With the increase of $Re$, the flow separation point moves upstream.

Figure 5(a) and 5(b) illustrate $\overline{L}_s$, $\overline{\theta}_s$ when wall velocity slips. Obviously, wall slip suppresses the appearance of downstream recirculation and make the wall separation point move towards rear stagnation point. When $Kn$ is more than $Kn^*$, the recirculation disappears completely, while $\overline{L}_s$ is zero and $\overline{\theta}_s$ is $180^0$. In addition, before the transitional $Kn^{**}$, $\overline{L}_s$ increases with the increase of $Kn$. This is because the flow stability becomes better with the increase of $Kn$ (Figure 3(b)).

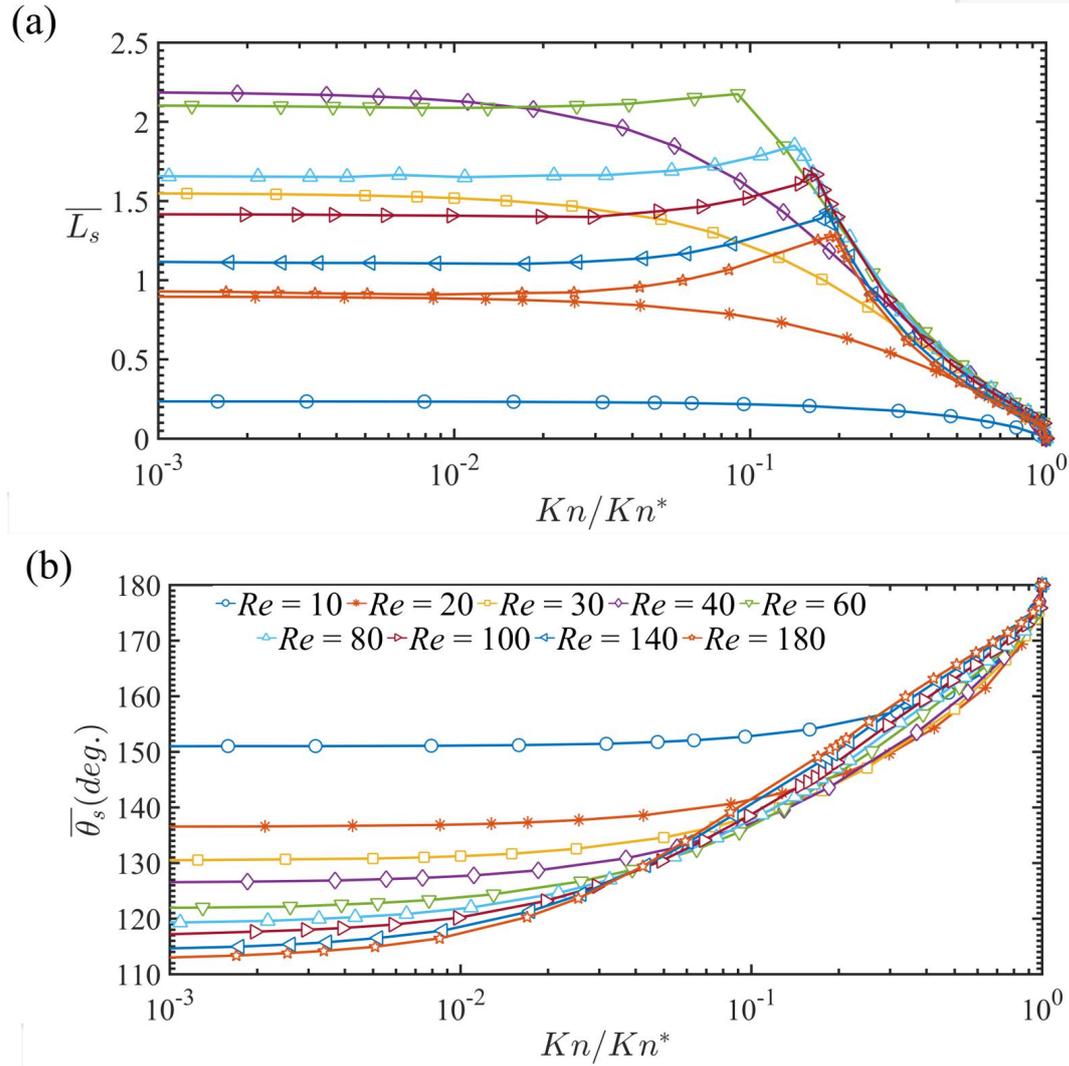

**Figure 5.** The relation between (a) $\overline{L}_s$, (b) $\overline{\theta}_s$ and $Kn$ for $Re$ = 10, 20, 30, 40, 60, 80, 100, 140 and 180. $Kn$ is normalized by $Kn^*$.

### 4.2 Velocity distribution on the cylinder surface

The velocity on the wall is not zero due to the wall slip. The velocity distribution at the wall surface is considered. Figure 6 presents the time-averaged wall tangential velocity distributions for $Re$ = 1, 40, 60, 140 considering different $Kn$. And it is redrawn in figure 7 in consideration of different *Re*. For a fixed *Re*, the flow is faster near the cylindrical wall as *Kn* increases, especially in front part of the cylinder as shown in figure 6. The points that the velocity reaches maximum with arrows are marked in figure 6, which move towards the tailing rear stagnation point. For a fixed *Kn*, the slip velocity on the wall becomes faster as *Re* increases, as shown in figure 7. If there exists an interval with $\overline{u}_\tau < 0$ on the wall, a flow separation and a recirculation bubble occurs in the wake field. As shown in figure 6 and figure 7, it is more difficult to generate negative velocity on the wall surface when the wall velocity slips. The recirculation bubble disappears completely when *Kn* is over than $Kn^*$.

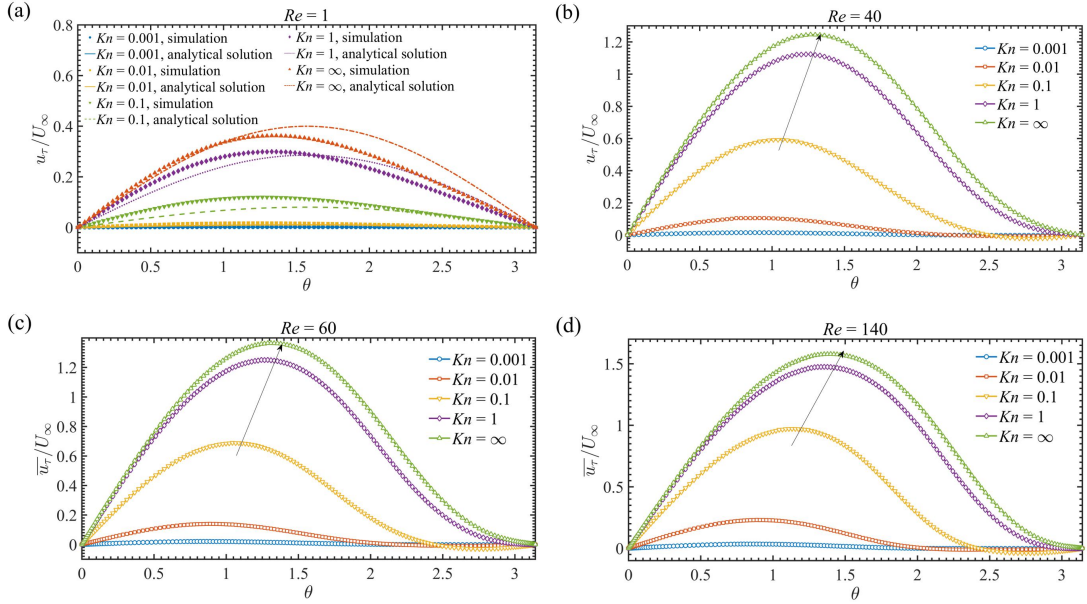

**Figure 6.** The time-averaged velocity profile on the wall for (a) *Re* = 1, (b) *Re* = 40, (c) *Re* = 60, and (d) *Re* = 140. In figure 6(a), velocity profile is compared with equation (12).

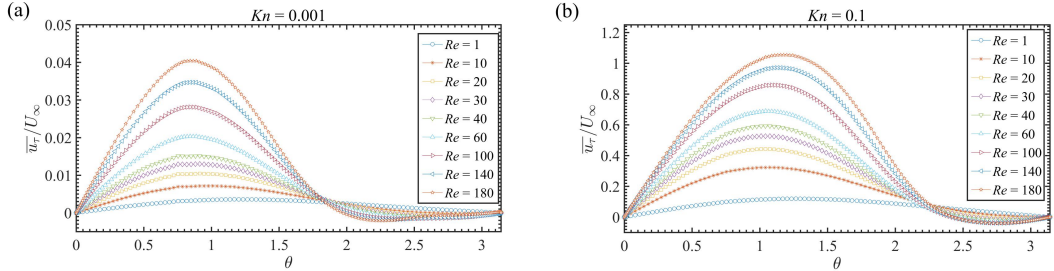

**Figure 7.** The time-averaged velocity profile on the wall for (a) *Kn* = 0.001, (b) *Kn* = 0.1.

The distribution of root mean square tangential velocity ($u_{rms}$) on cylindrical wall for different *Re* and *Kn* are shown in figure 8 and figure 9. For a fixed *Re*, such as *Re* = 60, as *Kn* increases from 0 to 0.01, $u_{rms}$ increases. Then, with the continuous increase of *Kn*, $u_{rms}$ decreases. Until *Kn* exceeds $Kn^{**}$, $u_{rms}$ becomes zero. For a fixed *Kn*, such as 0.001 and 0.1, as *Re* increases, the fluctuation of wall tangential velocity becomes larger. It is worth pointing out that when *Kn* is small, such as 0.001, there exists a high tangential velocity fluctuation on the cylindrical wall at a high *Re*, such as 180 ($u_{rms}/U_\infty > 0.04$).

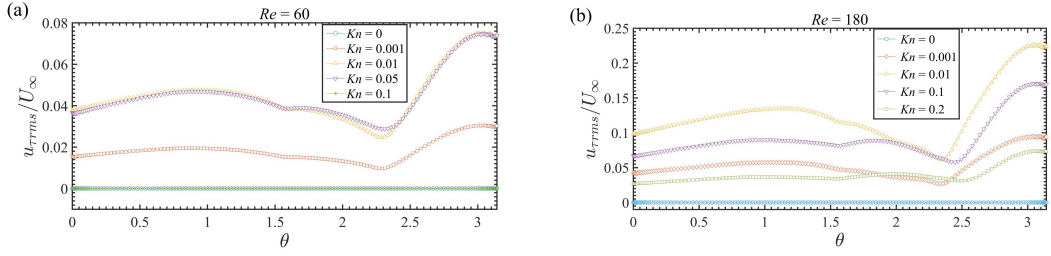

**Figure 8.** The root mean square velocity profile on the wall for (a) *Re* = 60, (b) *Re* = 180.

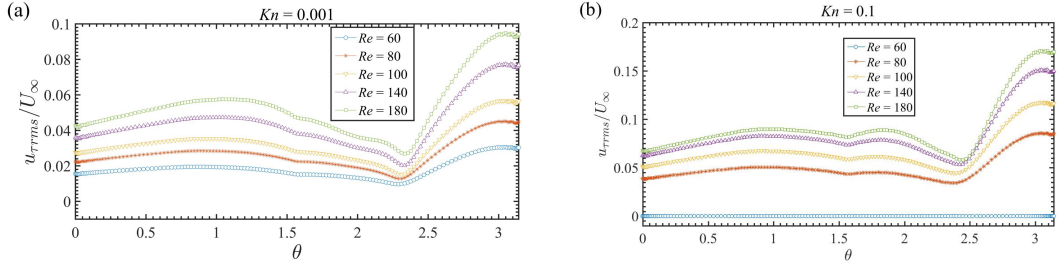

**Figure 9.** The root mean square velocity profile on the wall for (a) *Kn* = 0.001, (b) *Kn* = 0.1.

Li *et al.* [24] provided an analytical solution of tangential velocity on cylindrical wall as follows:

$$\frac{u_\tau}{U_\infty} = \frac{\dfrac{2Kn}{1+2Kn}}{-\ln\left(\dfrac{Re}{8}\right)+1-\gamma'-\dfrac{1}{2+4Kn}} \sin\theta \quad \text{when} \quad Re \ll 1, \quad (13)$$

where *γ'* is Euler constant. The theoretical solution is obtained by matched-asymptotic expansion. This equation is more accurate when *Re* is small. According to this equation, the velocity distributes front-back symmetry. We compare this equation with our numerical results for *Re* = 1 shown in figure 6(a). Obviously, the numerical solution is consistent with the analytical solution in order of magnitude, but the data difference is obvious. There is some asymmetry in the front and back of the velocity distribution on the cylinder due to the flow convection. From equation (13), the maximum tangential velocity is,

$$\frac{u_{\tau max}}{U_\infty} = \frac{\dfrac{2Kn}{1+2Kn}}{-\ln\left(\dfrac{Re}{8}\right)+1-\gamma'-\dfrac{1}{2+4Kn}} \quad \text{when} \quad Re \ll 1, \quad (14)$$

If *Kn* ≈ 0, The equation (13) can be reduced to

$$\frac{u_\tau}{U_\infty} \approx \frac{2Kn}{-\ln\left(\dfrac{Re}{8}\right)+\dfrac{1}{2}-\gamma'} \cdot \sin\theta \quad \text{when} \quad Re \ll 1 \text{ and } Kn \approx 0. \quad (15)$$

Because *γ'* is about 0.57722, if *Re* ≪ 1, −ln(*Re*/8) is far greater than *γ'* − 0.5. For example, when *Re* = 1, −ln(*Re*/8) is about 2.0794, while *γ'* − 0.5 is about 0.07722. From equation (15),

$$u_\tau/U_\infty \sim -1/\ln(Re/8) \quad \text{when} \quad Re \ll 1 \text{ and } Kn \approx 0. \tag{16}$$

At low $Re$, the velocity on the cylindrical wall is proportional to $-1/\ln(Re/8)$. The higher the $Re$, the faster the velocity on the wall surface. From equation (15),

$$u_\tau/U_\infty \sim Kn \quad \text{when} \quad Re \ll 1 \text{ and } Kn \approx 0. \tag{17}$$

When $Kn$ is small, the velocity on the cylindrical wall is proportional to $Kn$. This equation is applicable not only to low $Re$ ($Re \ll 1$) but also to high $Re$ (up to 180). Then, another theoretical analysis is given. From the definition formula of friction drag coefficient equation (6) and equation (11), we can get

$$\overline{C_{dv}} = 4\pi \frac{\overline{u_{avg}}/U_\infty}{Re \cdot Kn}, \tag{18}$$

where, $\overline{u_{avg}}$ is averaged $x$-velocity on cylindrical wall. $\overline{u_{avg}}$ is positively correlated with $\overline{u_{\tau max}}$. If $Kn$ is approach to 0, $\overline{u_{avg}}/U_\infty$ is near zero. From equation (18), we have

$$\overline{u_{avg}}/U_\infty \sim Kn, \tag{19}$$

$$\overline{u_{\tau max}}/U_\infty \sim Kn. \tag{20}$$

Variation of $\overline{u_{\tau max}}$ with $Kn$ ($Kn \leq 0.01$) for different $Re$ is shown in figure 11. For $Re \leq Re^{**}$, there exists a linear relationship between $\overline{u_{\tau max}}$ and $Kn$. However, when $Re$ is over $Re^{**}$, the linear interval is short. This may be due to the velocity fluctuation on the cylindrical wall surface discussed in figures 8 and 9. However, when $Kn$ is large, this linear relation is no longer applicable, which is shown in figure 12(a). $u_{\tau max}$ is a S-type increasing as $Kn$ (in logarithmic coordinate system) increases for a fixed $Re$.

For flow over no-slip circular cylinder, $\overline{C_{dv}}$ is linearly related with $Re$ in log-log coordinate system as shown in figure 10. This relation could be written as follows,

$$\left(\overline{C_{dv}}\right)_{Kn=0} \sim Re^{-0.6249} \quad \text{when} \quad 1 \leq Re < Re^{**}, \tag{21}$$

$$\left(\overline{C_{dv}}\right)_{Kn=0} \sim Re^{-0.5616} \quad \text{when} \quad Re^{**} \leq Re \leq 180. \tag{22}$$

Sen $et\ al.$ [36] gave the index is -0.60 when $6 \leq Re \leq 40$. When $Kn$ is small, $\overline{C_{dv}}$ changes a little, namely,

$$\left(\overline{C_{dv}}\right)_{Kn \approx 0} \approx \left(\overline{C_{dv}}\right)_{Kn=0}. \tag{23}$$

From equations (18), (21-23), a scaling law could be easily obtained as follows,

$$\overline{u_{avg}}/U_\infty \sim Re^{0.3751} \quad \text{when} \quad 1 \leq Re < Re^{**} \text{ and } Kn \approx 0, \tag{24}$$

$$\overline{u_{avg}}/U_\infty \sim Re^{0.5616} \quad \text{when} \quad Re^{**} \leq Re \leq 180 \text{ and } Kn \approx 0. \tag{25}$$

Variation of $\overline{u_{avg}}$ with $Re$ for different $Kn$ is shown in figure 12(c). The coordinate system in the figure is log-log coordinate system. $\log\left(\overline{u_{avg}}\right)$ is linear to $\log(Re)$ when $Kn$ is less than 0.03. In numerical simulation, the scaling indexes are 0.3747 and 0.5564, respectively, which are very close to equations (24) and (25). At $Re = \infty$ and $Kn = \infty$, the tangential velocity distribution on the wall satisfies the potential flow solution, which could written as follows [30]:

$$u_\tau = 2U_\infty \cdot \sin\theta, \quad \text{when} \quad Re = \infty \text{ and } Kn = \infty. \tag{26}$$

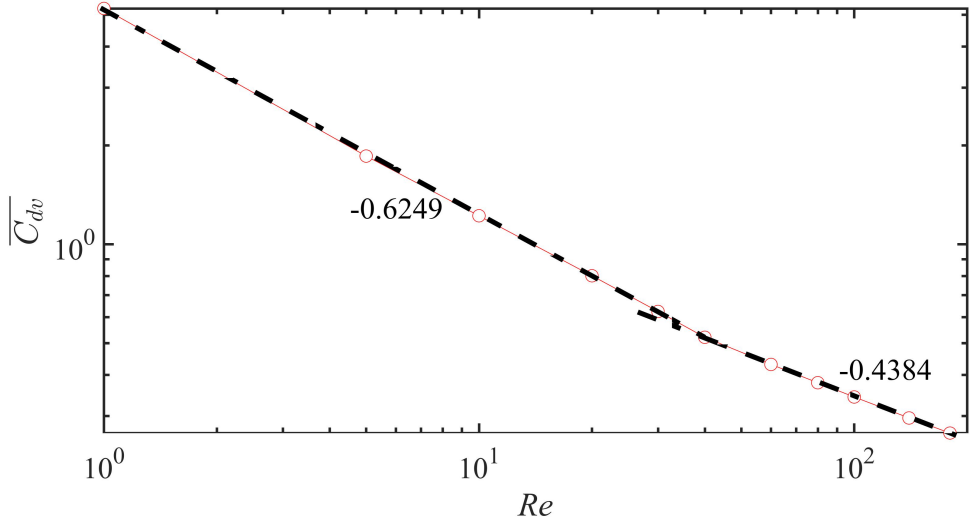

**Figure 10.** Variation of $\overline{C_{dv}}$ with $Re$ for no-slip circular cylinder. The dotted line in the figure is an auxiliary line, indicating that it is a straight line.

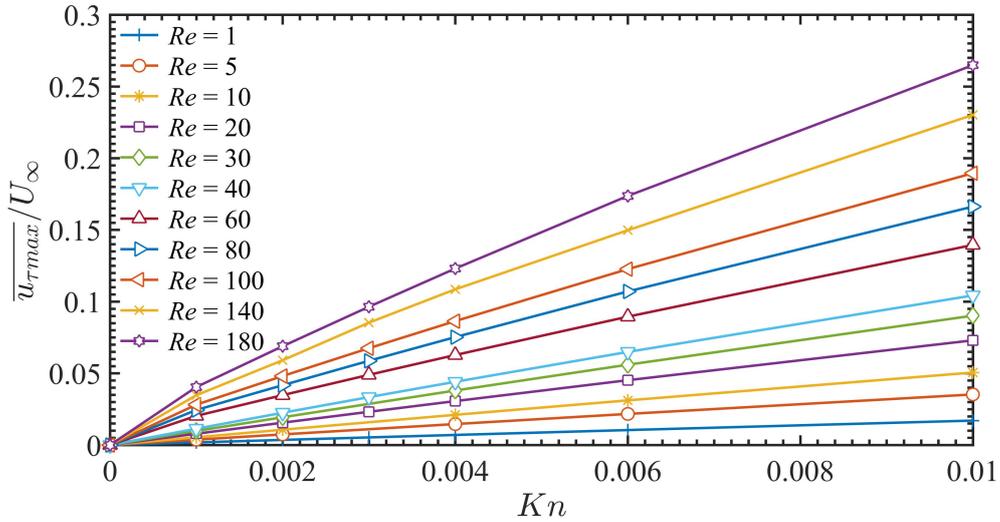

**Figure 11.** $\overline{u_{\tau max}}$ on the cylinder wall for different $Re$ and $Kn$ for $Kn \leq 0.01$.

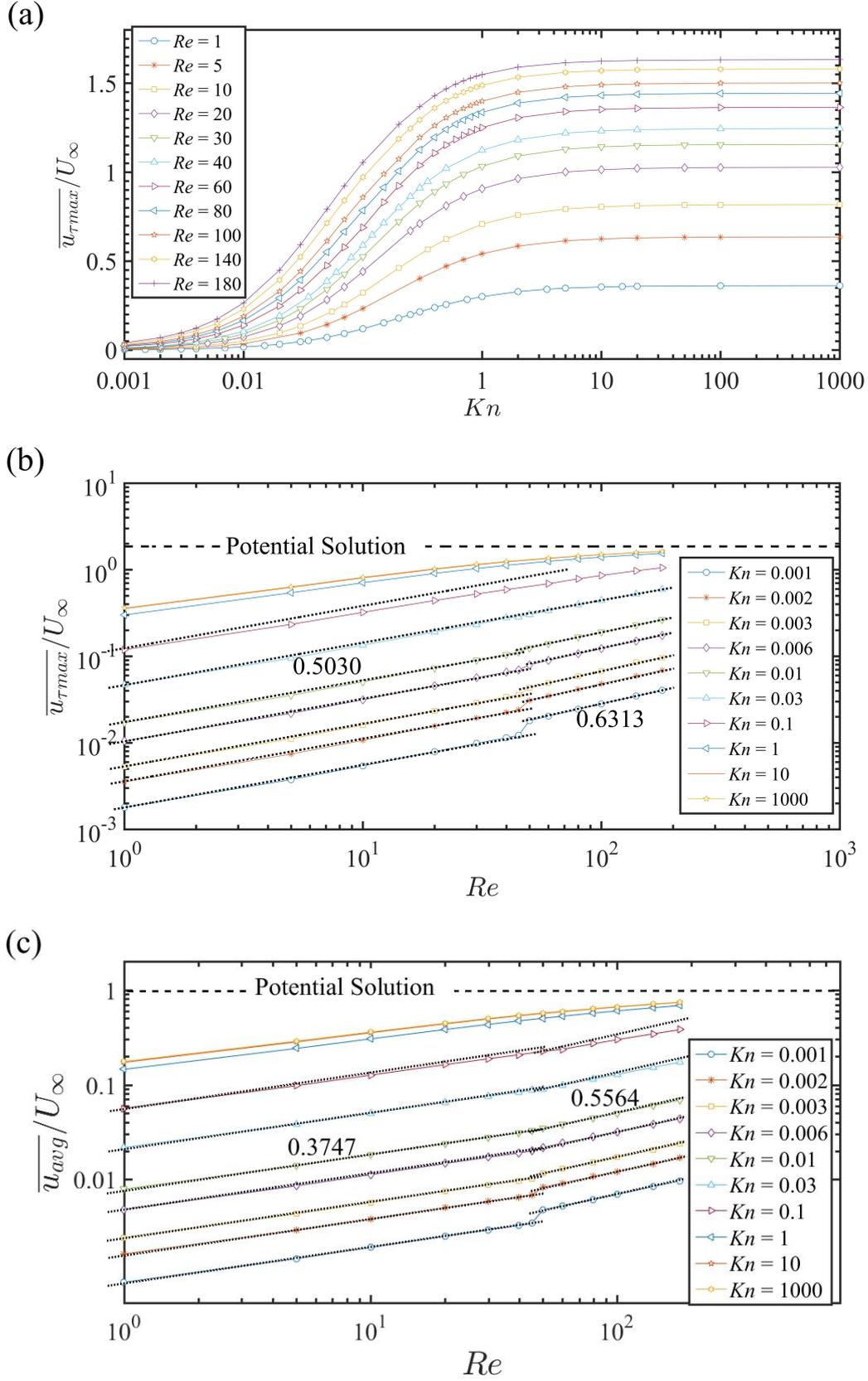

**Figure 12.** (a-b) $\overline{u_{\tau max}}$ on the cylinder wall for different $Re$ and $Kn$ and (c) $\overline{u_{avg}}$ for different $Re$ and $Kn$. The dotted line in the figure is an auxiliary line, indicating that it is a straight line.

From the equation (26), it is easy to get:

$$u_{avg} = U_\infty, \quad \text{when} \quad Re = \infty \text{ and } Kn = \infty. \tag{27}$$

$u_{avg} = U_\infty$ is an asymptotic line for high $Re$ and high $Kn$ plotted as along dotted line in figure 12(c).

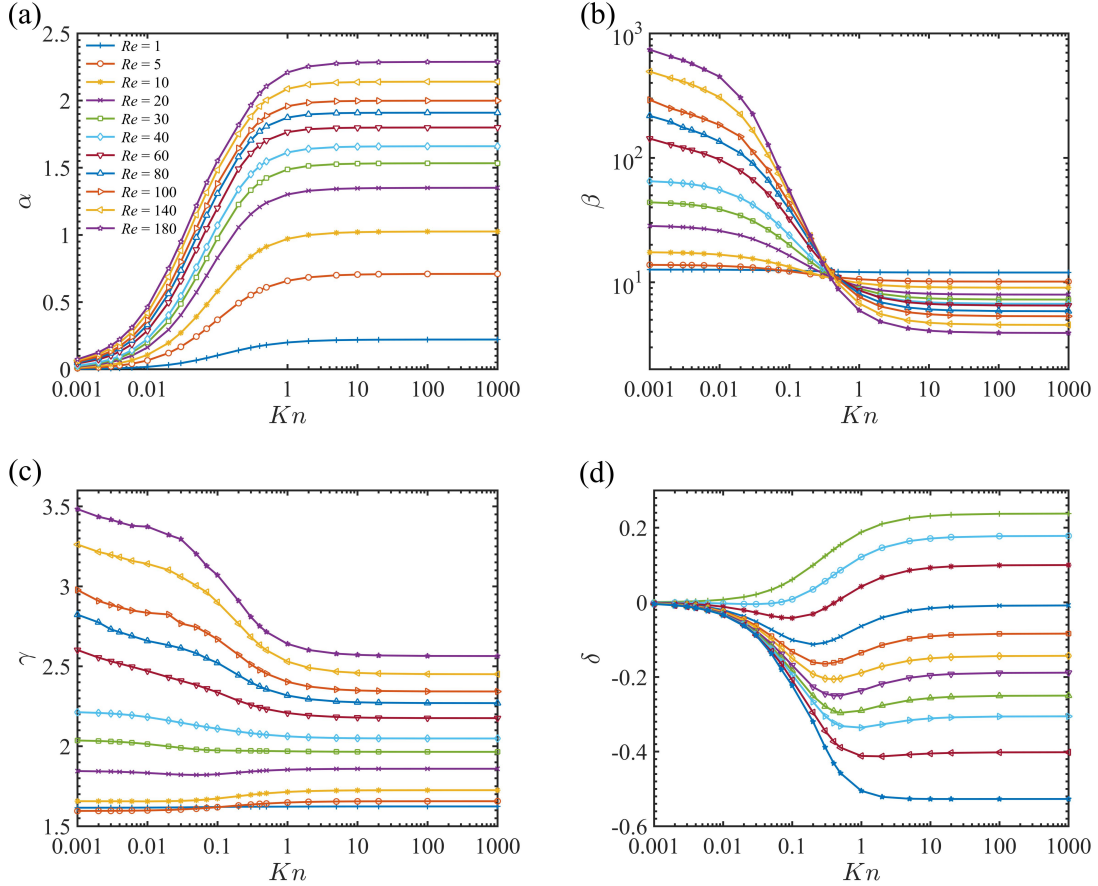

**Figure 13.** The coefficients for (a) $\alpha$, (b) $\beta$, (c) $\gamma$ and (d) $\delta$.

Similar to $\overline{u_{avg}}$, a scaling law could also be applied to $\overline{u_{\tau max}}$ written as follows,

$$\overline{u_{\tau max}}/U_\infty \sim Re^{0.5030} \quad \text{when} \quad 1 \leq Re < Re^{**} \text{ and } Kn \approx 0, \tag{28}$$

$$\overline{u_{avg}}/U_\infty \sim Re^{0.6313} \quad \text{when} \quad Re^{**} \leq Re \leq 180 \text{ and } Kn \approx 0. \tag{29}$$

This scaling law is obtained through numerical simulation as shown in figure 12(b). $u_{\tau max} = 2U_\infty$ (potential solution) is an asymptotic line for high $Re$ and high $Kn$ plotted as along dotted line in figure 12(b).

When $Re>1$, the equation (13) is not suitable to describe the velocity on the cylinder surface. Therefore, another formula is used to describe the relationship between wall tangential velocity $\overline{u_\tau}$ and the angle $\theta$,

$$\frac{\overline{u_\tau}}{U_\infty} = \left[\frac{\alpha}{1+\beta \cdot e^{-\gamma(\pi-\theta)}} + \delta\right]\sin\theta. \tag{30}$$

The coefficients for $\alpha$, $\beta$, $\gamma$ and $\delta$ are functions to $Re$ and $Kn$, where $\alpha$ and $\delta$ have the dimension of velocity. $\delta$ is an independent number and $\alpha$ is number, however, multiplied with a complicated formula. $\beta$ and $\gamma$ are two dimensionless coefficients. The coefficients for different $Re$ and $Kn$ are plotted in figure 13. Figure 14 compares the calculated wall slip velocity and the fitting results, which are fitted well to our simulated results. This formula may be used for theoretical analysis in the future.

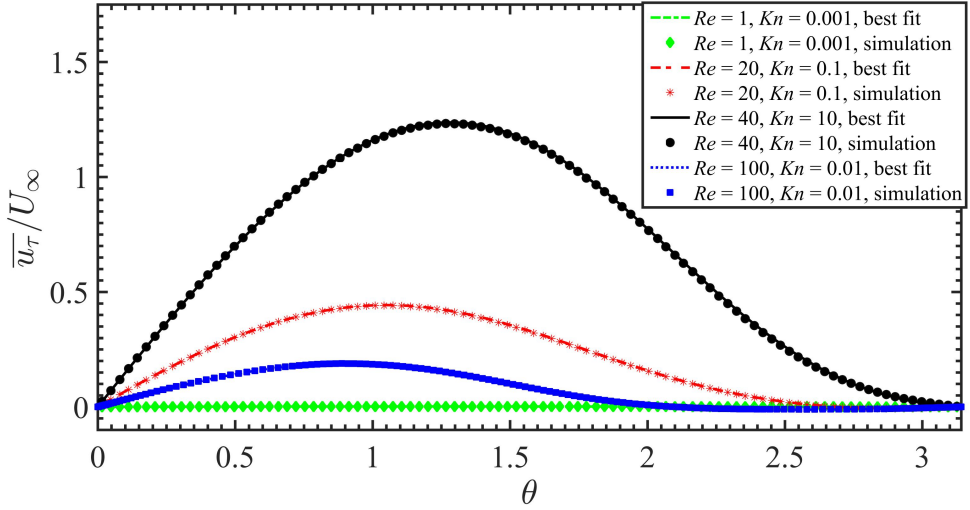

**Figure 14.** The calculation results and best fitting tangential velocity distribution comparison for $(Re, Kn) = (1, 0.001), (20, 0.1), (40, 10)$ and $(100, 0.01)$.

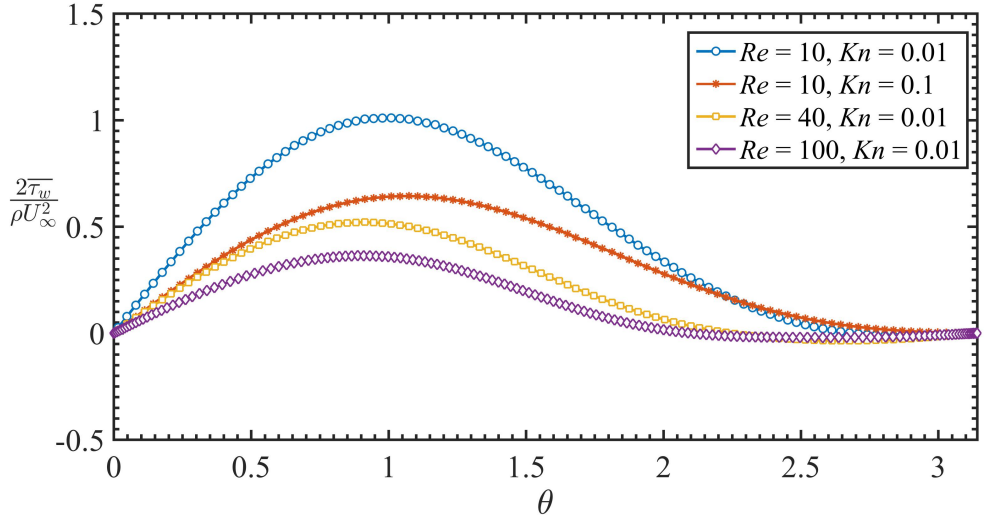

**Figure 15.** The relation between wall shear stress coefficient and angle ($\theta$) for $(Re, Kn) = (10, 0.01), (10, 0.1), (40, 0.01)$ and $(100, 0.01)$.

**4.3 Distribution of shear stress and pressure in the cylinder wall.**

Figure 15 shows the time-averaged shear stress distribution on the cylindrical wall. The cases parameters are chosen as (*Re*, *Kn*) = (10, 0.01), (10, 0.1), (40, 0.01) and (100, 0.01). For a fixed *Re*, as *Kn* increases, the shear stress becomes lower for wall slip. For a fixed *Kn*, as *Re* increases, the shear stress also becomes lower for the decrease of viscosity near the cylindrical wall.

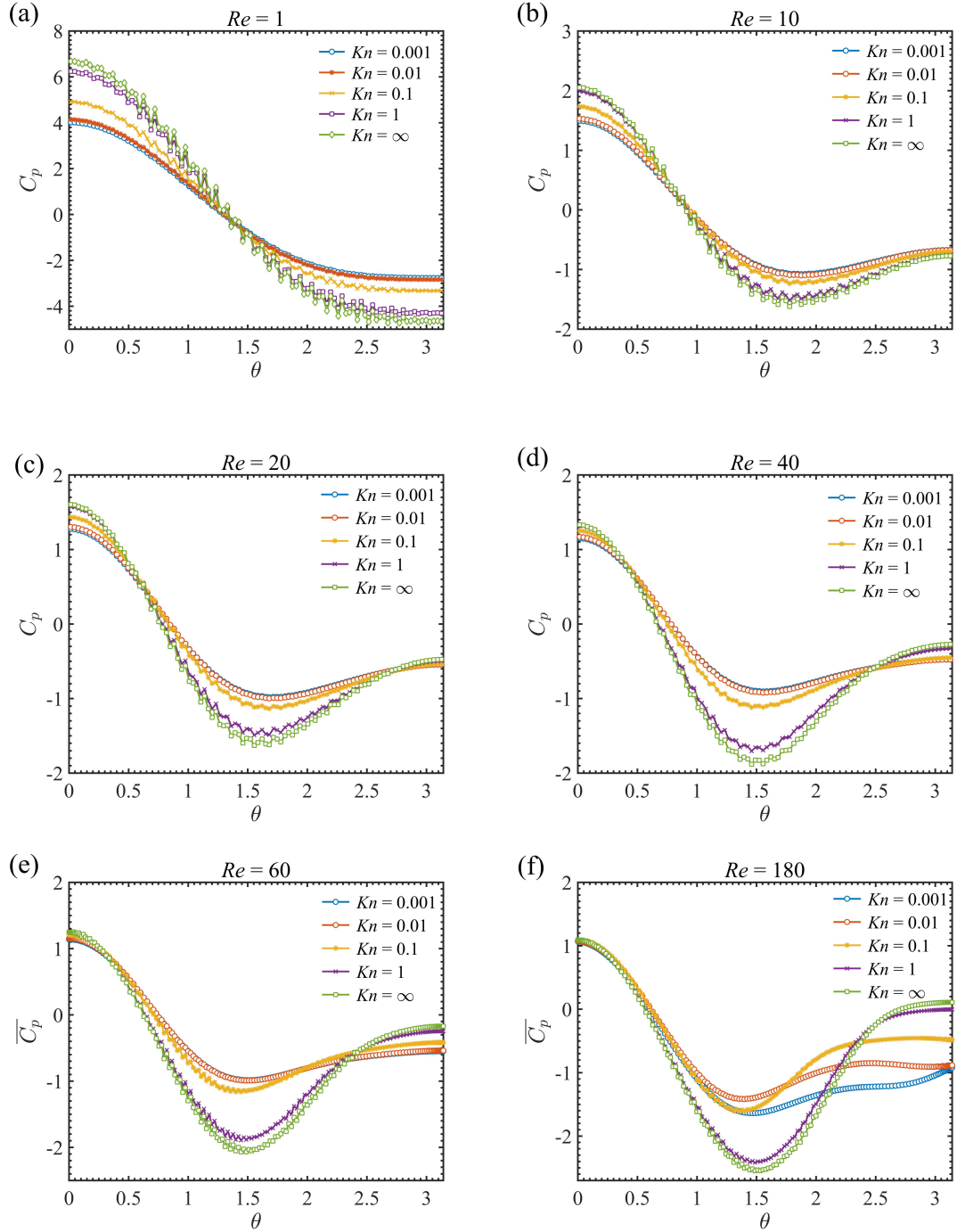

**Figure 16.** The time-averaged pressure coefficient on the wall for (a) *Re* = 1, (b) *Re* = 10, (c) *Re* = 20, (d) *Re* = 40, (e) *Re* = 60 and (f) *Re* = 180.

The time-averaged pressure coefficient ($\overline{C_p}$) along with angle $\theta$ for selected *Re* and *Kn* are shown in figure 16. The maximum wall pressure coefficient ($\overline{C_{pmax}}$) all appears at the front end of the cylinder. The minimum wall pressure coefficient ($\overline{C_{pmin}}$) all appears at around $\theta = \pi/2$.

$\overline{C_{pmax}}$ for different $Re$ and $Kn$ are shown in figure 17(a). $\overline{C_{pmax}}$ increases with the increase of $Kn$, especially for small $Re$, such as $Re = 1$. For $Re = 1$, $\overline{C_{pmax}}$ is 3.976 ($Kn = 0$), while 6.676 ($Kn = \infty$). For small $Re$, wall shear stress is large for high viscosity near cylinder's wall. With the increase of $Kn$, wall stress decreases, which leads to the pressure increase at the front end of the cylinder.

The difference between the maximum wall pressure coefficient and the minimum wall pressure coefficient ($\overline{C_{pmax}} - \overline{C_{pmin}}$) for different $Re$ and $Kn$ are shown in figure 17(b). For a fixed $Re$, a greater pressure difference on the wall for wall slip. For example, for $Re = 1$, $\overline{C_{pmax}} - \overline{C_{pmin}}$ = 6.796 ($Kn = 0.001$) and 8.262 ($Kn = 0.1$). For all, wall slip enables the pressure on the front part of cylinder's wall to increase, especially for small $Re$, and leads to wall maximum pressure coefficient difference ($\overline{C_{pmax}} - \overline{C_{pmin}}$) increase for all $Re$. The large pressure difference and lower shear stress on the wall enhance the energy driving on the wall, resulting in a higher flow velocity on the wall.

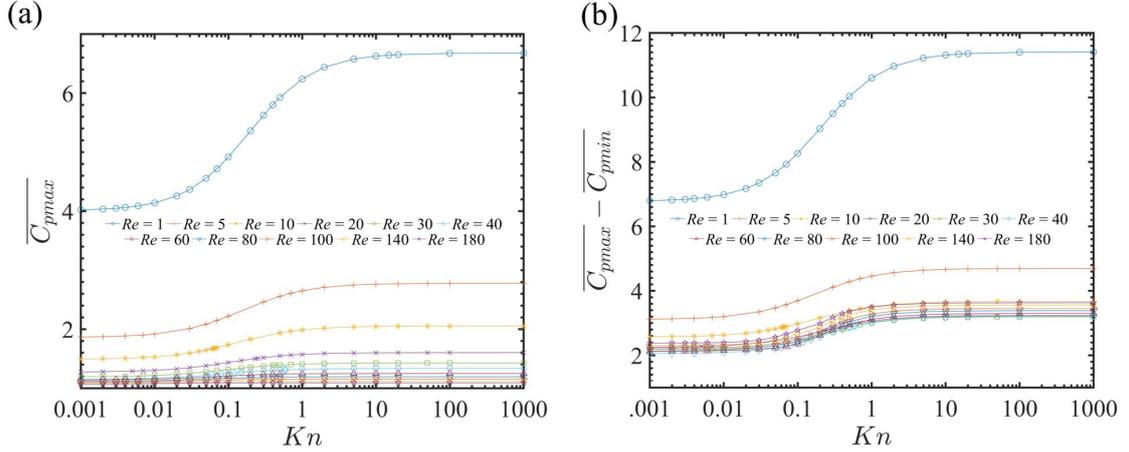

**Figure 17.** (a) Time-averaged maximum wall pressure coefficient and (b) difference between time-averaged maximum and minimum wall pressure coefficients for different $Re$ and $Kn$.

Another phenomenon need to be pointed out that $\overline{C_p}$ at $\theta = \pi$ becomes larger when wall velocity slips. For example, for $Re = 180$, $\overline{C_p}$ at $\theta = \pi$ is -0.948 ($Kn = 0$) and -0.4774 ($Kn = 0.1$). This is due to wake flow stability enhancement (as shown in figure 3(b)) when wall velocity slips.

**4.4 Drag reduction**

Figures 18(a) and 18(b) show the variation of the total drag reduction ($DR = 1 - \frac{\overline{C_d}}{\overline{C_{d(Kn=0)}}}$) for various $Kn$ and $Re$. Wall slip reduces the drag acting on the cylinder. $DR$ increases as $Kn$ increases with S-type when $Re$ is fixed, as shown in figure 18(a). For a fixed $Kn$, $DR$ is linear to $Re$ ($Re \leq Re^{**}$) in the log-log coordinates, namely,

$$\log(DR) \sim \log(Re) \quad \text{when} \quad Re \leq Re^{**} \text{ and } Kn \leq 0.1 . \tag{31}$$

When $Re$ is about $Re^{**}$, a turning point appears in the figure 18(b). $DR$ has an approximate linear relationship with $Re$ ($Re > Re^{**}$) in the log-log coordinates, while the slope is obviously higher than it when $Re \leq Re^{**}$.

Figure 19 shows the relationship between $DR$ and $Kn$ for $Kn \leq 0.05$. $DR$ is almost linearly to $Kn$ at this range in the log-log coordinate, namely,

$$\log(DR) \sim \log(Kn) \quad \text{when} \quad Kn \leq 0.05. \tag{32}$$

This equation is more precisely when the $Re$ is smaller and $Kn$ is smaller.

Drag includes two parts, differential pressure drag and friction drag. Differential pressure drag reduction ratio ($DR_p = 1 - \frac{\overline{C_{dp}}}{\overline{C_{dp}}_{(Kn=0)}}$) and friction drag reduction ratio ($DR_v = 1 - \frac{\overline{C_{dv}}}{\overline{C_{dv}}_{(Kn=0)}}$) with changing $Kn$ are shown in figure 20(a) and 20(b), respectively. When $Re$ is less than 20, $DR_p$ increases as $Kn$ increases. This result is quite out of the ordinary. However, Maghsoudi et al. [31] also obtaied the similar results. Conversely, if $Re$ is greater than 20, $DR_p$ decreases as $Kn$ increases. The friction drag acting on cylindrical wall decreases due to wall slip. Moreover, the lines of $DR_v \sim Kn$ in figure 20(b) are almost overlapped for different $Re$. That is, $DR_v$ is found only a function of $Kn$ and almost independent with $Re$.

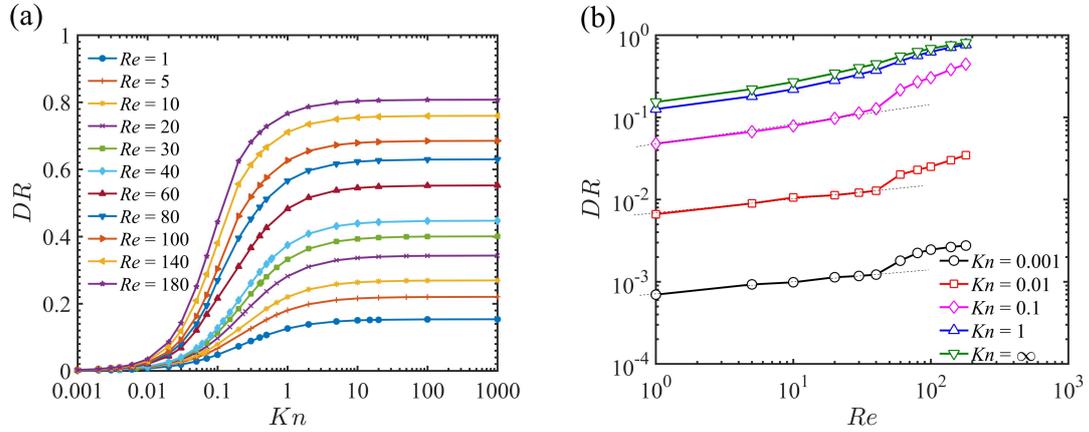

**Figure 18.** Total drag reduction ($DR$) for different $Re$ and $Kn$. The dotted line in the figure is an auxiliary line, indicating that it is a straight line.

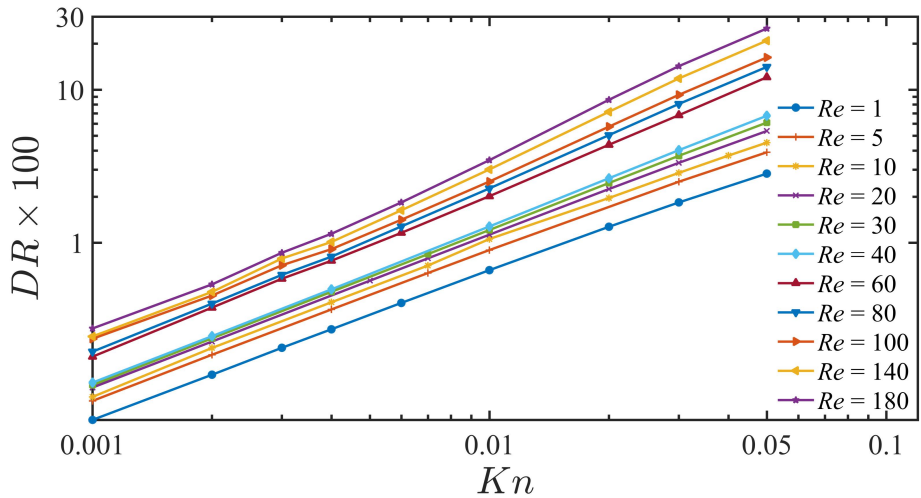

**Figure 19.** Variation of $DR$ for small $Kn$.

Figure 21 demonstrates variation diagram of $f(Re, Kn) = \frac{\overline{C_{dp}} - \overline{C_{dp}}_{(Kn=0)}}{\overline{C_{dv}} - \overline{C_{dv}}_{(Kn=0)}}$, where $f$ is used to describe total drag reduction contributed from differential pressure drag reduction ($DR_p$) and friction drag reduction ($DR_v$). When $f < 0$, the differential pressure drag increases compared to no slip boundary. When $f > 1$, the contribution of differential pressure drag reduction to the reduction of total drag is higher than that of friction drag. The short dash line delegates $f = 0$ and the long dash line delegates $f = 1$. Below the short dashed line, wall slip increases the pressure drag compared to no-slip wall, which appears in the range of small $Re$ ($Re \leq 20$). $f < 0$ also could appears at low $Kn$ for high $Re$ ($Re > 20$). At this range, the contribution of differential pressure drag ($DR_p$) to the total drag reduction ($DR$) is negative. Above the long dashed line in the figure, $Re > 60$ and $Kn$ over a critical $Kn$, the differential pressure drag ($DR_p$) contributes more to the total drag reduction ($DR$).

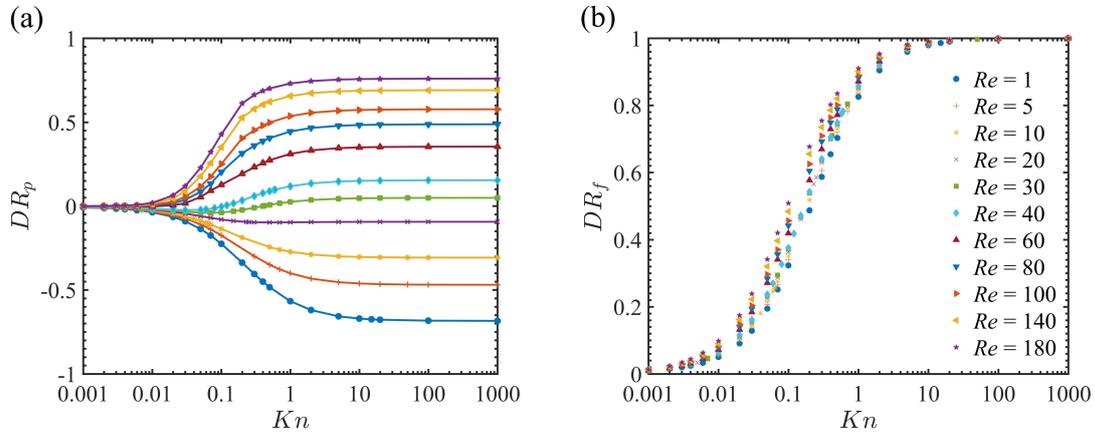

**Figure 20.** Variation of (a) $DR_p$ and (b) $DR_v$ for the different $Re$ and $Kn$.

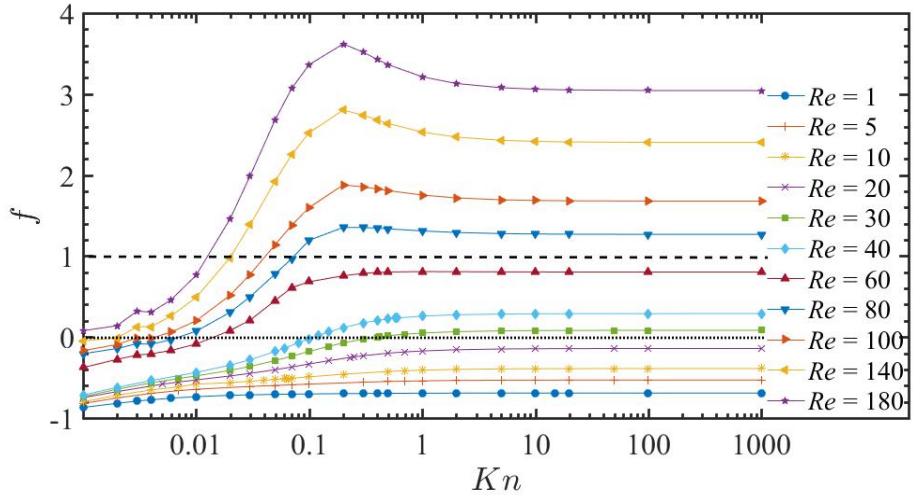

**Figure 21.** Contribution function of drag reduction $f(Re, Kn)$.

## 5. Conclusion remarks

Flow over a circular cylinder with slip wall for *Re* from 1 to 180 are studied through numerical simulation. Slip effect is considered as a simple slip length model. Correspondingly, Knudsen number (*Kn*) is a dimensionless number to evaluate slip length. *Kn* is considered from 0 to ∞. The commercial software Fluent is used to solve flow control and the slip boundary equations. The time-averaged recirculation length and separation angle, the slip velocity distribution on the wall, wall shear stress distribution, wall pressure distribution, drag coefficient and drag reduction are investigated.

A greater velocity distributes on cylindrical wall when wall velocity slips, especially on front part of cylindrical wall. The separation points on the wall moves toward the rear stagnation point, so that it has shorter wake bubble length than that of no-slip condition. The tailing vortex disappears completely if the two points coincide. A formula $\overline{u_\tau} = U_\infty \cdot \left[ \dfrac{\alpha}{1+\beta e^{-\gamma(\pi-\theta)}} + \delta \right] \sin\theta$ is given to describe the distribution of tangential velocity on the wall, where the coefficients (*α*, *β*, *γ*, *δ*) are related to *Re* and *Kn*, and $U_\infty$ is the incoming velocity. A theoretical analysis indicates the maximum tangential velocity on the cylinder wall ($\overline{u_{\tau max}}$) and *Re*, *Kn* satisfy such two scaling laws, $\log(\overline{u_{\tau max}}) \sim \log(Re)$ and $\overline{u_{\tau max}} \sim Kn$ when *Kn* is low, which is proved to be true from numerical simulation.

A greater pressure coefficient difference ($\overline{C_{pmax}}-\overline{C_{pmin}}$) and lower shear stress occurs on the cylinder wall when the cylinder's wall velocity slips, which provides the driving energy of the wall flow velocity. When wall velocity slips, the pressure at the tailing end of the cylinder increases for wall flow stability enhancement, especially when *Re* is high, such as, *Re* = 180.

The drag reduction for wall slip introduction is specified. At low *Kn*, drag reduction ratio (*DR*) is found linear to *Kn* in log-log logarithmic coordinate system when *Kn*≤0.05. The relation between *DR* and *Re* satisfies log(*DR*)~log(*Re*) when *Re* is lower than $Re^{**}$ (the critical *Re* for wake vortex shedding). At low *Re*, $DR_v$ (the friction drag reduction) is the main source of *DR* while $DR_p$ (the pressure drag reduction) makes a negative contribution to the total *DR*. However, $DR_p$ (the differential pressure drag reduction) contributes the most to *DR* at high *Re* (*Re*>~60) and *Kn* over a critical number. $DR_v$ is found only a function of *Kn* and almost independent with *Re*.

Our numerical results could be regarded as a supplement of these extreme conditions of Li *et al.* [24] *(Re* <<1) and Kumar *et al.* [30] (*Re*→ ∞ and *Kn*→ ∞) at moderate *Re* and *Kn*.

## 6. Data Availability Statement

The data that support the findings of this study are available from the corresponding author upon reasonable request.

## 7. Acknowledgement

The author Sai Peng would like to thank the financial support from the National Natural Science Foundation of China (NSFC, Grant No. 12002148), Guangdong Basic and Applied Basic Research Foundation (Grant No. 2022A1515011057).